\documentclass[aps,prb,superscriptaddress,twocolumn,longbibliography]{revtex4-2}
\usepackage{bbm}
\usepackage{graphicx}
\usepackage{dcolumn}
\usepackage{bm}
\usepackage{subfigure}
\usepackage{amsmath}
\usepackage{amssymb}

\usepackage{hyperref}
\hypersetup{
	colorlinks = true,
	citecolor  = magenta,  
	linkcolor  = magenta 
}

\usepackage{attachfile}

\newcommand{\bk}{\boldsymbol k}

\usepackage{times}

\begin{document}
\title{Topological superconductivity in layered $d$-wave altermagnetic metals}

\author{Yijie Mo}
\email{moyj36@mail.sysu.edu.cn}
\affiliation{Guangdong Provincial Key Laboratory of Magnetoelectric Physics and Devices, State Key Laboratory of Optoelectronic Materials and Technologies, School of Physics, Sun Yat-sen University, Guangzhou 510275, China}

\author{Zhongbo Yan}
\email{yanzhb5@mail.sysu.edu.cn}
\affiliation{Guangdong Provincial Key Laboratory of Magnetoelectric Physics and Devices, State Key Laboratory of Optoelectronic Materials and Technologies, School of Physics, Sun Yat-sen University, Guangzhou 510275, China}

\date{\today}

\begin{abstract}
We investigate multilayer $d$-wave altermagnetic metals with odd-parity helical $p$-wave pairing. In the minimal bilayer system, interlayer tunneling reconstructs the spin-split Fermi surfaces and generates two layer-hybridized sectors with distinct effective chemical potentials. Without Rashba spin-orbit coupling (RSOC), the emergent mirror symmetry and subsystem chiral symmetry stabilize phases with different numbers of helical Majorana edge-mode pairs and additional counterpropagating pairs, jointly characterized by the mirror Chern number and the winding numbers defined along high-symmetry lines. The combination of RSOC and the altermagnetic exchange field gaps out the helical Majorana pairs while, in a certain parameter regime, leaving another pair of counterpropagating Majorana edge modes gapless, giving rise to second-order topological phases with isolated or twofold Majorana corner modes, as well as a hybrid-order phase in which gapless Majorana edge modes coexist with Majorana corner modes. We identify a layer symmetry that elucidates the physical origin of these phases. For an arbitrary $N$-layer system, such a symmetry naturally arises from the interlayer-coupling matrix, allowing $N$-fold Majorana corner modes to emerge when all layer-hybridized sectors enter the nontrivial second-order regime. Our results establish the interplay among the layer degree of freedom, odd-parity pairing, and Fermi-surface configurations as a controllable route to engineering topological superconductivity in layered altermagnetic systems. 
\end{abstract}

\maketitle

\section{Introduction \label{sec1}}

Altermagnetism has recently emerged as a distinct class of collinear magnetism beyond conventional ferromagnetism and 
antiferromagnetism~\cite{smejkal2022beyond,smejkal2022emerging,wu2007,Hayami2019momentum,Hayami2020bottom,smejkal2020crystal,
Yuan2020giant,Yuan2021prediction,Shao20212021,Ma2021multifunctional}. Similar to ordinary collinear antiferromagnets, altermagnets possess fully compensated magnetic moments and hence vanishing net magnetization. 
However, unlike conventional antiferromagnets, where opposite-moment sublattices are related by translation or 
spatial inversion, those in altermagnets are connected via crystal rotations or mirror operations~\cite{Liu2022spin,Chen2024enumeration,Jiang2024enumeration,Xiao2024spin}. Consequently, their electronic bands can exhibit sizable nonrelativistic spin splitting even in the absence of spin-orbit coupling. 
A key feature of spin splitting in altermagnets is its symmetry pattern in momentum space. Notably, these patterns are remarkably rich and have led to two primary classification schemes. One coarse-grained framework is based on the parity of the spin splitting under momentum reversal, yielding three classes: even~\cite{smejkal2022beyond,smejkal2022emerging}, odd~\cite{Liu2026light,Huang2026light,Zhu2026floquet,Li2026floquet,Lin2026odd,Zeng2026odd,Zhuang2026odd,Luo2025}, and mixed parity~\cite{Zhuang2026mixed,Yue2026}. On the other hand, since opposite-moment sublattices are related by rotation or mirror symmetries, 
the spin-splitting patterns are also classified in terms of angular momentum harmonics~\cite{smejkal2022beyond,smejkal2022emerging} (e.g., $p$-wave, $d$-wave,
$f$-wave, $g$-wave) or mirror-based partitions~\cite{Zhuang2026stripe}. To date, numerous materials have been experimentally identified as 
altermagnets, such as $\mathrm{MnTe}$~\cite{Osumi2024observation,lee2024broken,Krempasky2024altermagnetic,Hajlaoui2024temperature},  $\mathrm{CrSb}$~\cite{Reimers2024direct,Ding2024large,Yang2025three,Zeng2024observation,Li2025topological,Lu2025signature},
$\mathrm{KV_{2}Se_{2}O}$~\cite{Jiang2025metallic,Wang2025atomic,Yang2026visualizing}, $\mathrm{CoNb_{4}Se_{8}}$~\cite{Regmi2025altermagnetism,Dale2026relativistic}, and $\mathrm{Rb_{1-\delta}V_{2}Te_{2}O}$~\cite{Zhang2025crystal}, 
to name a few. All materials reported to date belong to the even-parity class, while odd- and mixed-parity altermagnets remain experimentally elusive. A key challenge for their realization is that spinless time-reversal symmetry must be broken, either spontaneously through loop-current~\cite{Lin2026odd,Zeng2026odd,Leeb2026} or complex orbital order~\cite{Zhuang2026odd}, or dynamically through external driving~\cite{Liu2026light,Huang2026light,Zhu2026floquet,Li2026floquet}.

The momentum-dependent spin splitting of altermagnets has profound consequences for superconductivity~\cite{Zhu2023topological,Brekke2023two,Kristian2024many,Bose2024altermagnetism,
Carvalho2024unconventional,Parthenios2025spin,Maeda2025classification,Hong2025unconventional,
Wu2025intra,Fukaya2025superconducting,Parshukov2025exotic,Fu2026Floquet,Fu2026light,
Yokoyama2025Floquet,Zhang2024finite,Liu2026altermagnetism,Monkman2026,Zou2026AMSC,Lie2026,Lu2025AMSC}. We focus on even-parity altermagnets, which already have concrete material realizations. 
In an even-parity altermagnet, the even-parity spin splitting implies that the energy dispersion satisfies $E_{\sigma}(\bk)=E_{\sigma}(-\bk)$ ($\sigma=\{\uparrow,\downarrow\}$), but $E_{\sigma}(\bk)\neq E_{\bar{\sigma}}(\bk)$ for a generic momentum $\bk$. 
Similar to a conventional Zeeman field, the spin splitting in an even-parity altermagnet disfavors spin-singlet pairing with zero center-of-mass momentum, but can favor spin-singlet pairing with a finite center-of-mass momentum [Fulde–Ferrell–Larkin–Ovchinnikov (FFLO) states] 
under appropriate conditions~\cite{Chakraborty2024,Hu2025FFLO,Liu2026FFLO}, as it lifts the Kramers degeneracy between $E_{\sigma}(\bk)$ and $E_{\bar{\sigma}}(-\bk)$ enforced by time-reversal symmetry. Notably, the spin-split Fermi surfaces are perfectly compatible 
with spin-triplet pairing with zero center-of-mass momentum, since each electron on the Fermi surface can find 
another electron with the same spin but opposite momentum to form a Cooper pair. Within the framework of weak-coupling Bardeen-Cooper-Schrieffer (BCS) theory, such perfect nesting would lead to a pairing instability that can fully gap out the Fermi surface if an attractive interaction appears in the equal-spin pairing channel, even though the interaction is arbitrarily weak.
Indeed, numerous theoretical studies based on phenomenological or microscopic interactions have demonstrated that superconductivity with 
pure or dominated spin-triplet pairing can emerge in altermagnetic metals~\cite{Zhu2023topological,Brekke2023two,Kristian2024many,
Hong2025unconventional,Wu2025intra,Monkman2026,Zou2026AMSC}. 

In the study of topological superconductivity, odd-parity spin-triplet pairing is of particular interest, as it provides one of the most direct routes to realizing various topological superconducting phases, ranging from first-order topological superconducting phases~\cite{Read2000paired,Kitaev2001unpaired,Qi2009time,Kallin2016chiral,Zhang2008px+ipy,Fu2010odd,Sato2010topological,
Sato2009topological,Mo2025topological,Feng2022time, Qi2011topological} 
and higher-order topological superconducting phases~\cite{Zhu2018tunable,Wang2018weak,Wu2019higher,Yan2019higher,Kheirkhah2020,
Ahn2020Higher,Hsu2020inversion,Roy2020higher,Jahin2023higher,Chatterjee2024second,Bonetti2024interaction,
Amundsen2022controlling} to hybrid-order topological superconducting phases~\cite{Wu2023hybrid,Arouca2024mixed}.
Notably, in the weak-pairing limit, the topology of a fully gapped odd-parity superconductor is merely dictated by the Fermi-surface configuration in the normal state \cite{Fu2010odd,Sato2010topological,Sato2009topological,Yan2019higher}, which provides a simple 
guiding principle for realizing and designing topological superconductivity.

Thus far, most existing studies of topological superconductivity in altermagnet metals have focused on strictly two-dimensional monolayer models \cite{Zhu2023topological,Wu2025intra,Li2023MCM,Ghorashi2024altermagnetic,Zhu2024dislocation,Chatterjee2025interplay,Alam2026proximity,Li2024Realizing,
Yang2026topological,Zhu2026altermagnetic,Wan2026chiral,Liudeng2026}. Remarkably, two experimental groups have recently reported the observation of superconducting signatures 
in quasi-two-dimensional layered vanadium oxychalcogenides Na$_{2-x}$V$_{2}$Se$_2$O~\cite{Sun2026emergent} and CsV$_2$Se$_2$O~\cite{Li2026SC}.
Motivated by these experimental discoveries, Franz investigated the pairing instabilities in this family of layered altermagnets and identified spin-triplet superconductivity as the dominant pairing instability~\cite{Franz2026unconventional}. In particular, the leading pairing component remains an in-plane spin-triplet $p$-wave component, similar to the situation in the monolayer limit. These works suggest that layered altermagnets are promising platforms for hosting topological superconductivity.
However, the role of the layer degree of freedom in shaping topological superconductivity, as well as its evolution with increasing layer number, remains largely unexplored. This issue is particularly relevant in layered systems, where interlayer tunneling can generate bonding and antibonding bands with distinct effective chemical potentials and reconstruct the Fermi surfaces. As noted earlier, the topology of an odd-parity superconductor is closely tied to the Fermi-surface configuration. Therefore, the layer degree of freedom is expected to serve as an effective knob 
for tuning the topological superconducting phases~\cite{Nakosai2012topological,Mo2025topological,Wang2026intrinsic,Feng2022time,Mo2025coexistence,Mo2026layer}.

In this work, we investigate topological superconductivity in multilayer $d$-wave altermagnetic metals with spin-triplet helical $p$-wave 
pairing~\cite{Zhu2023topological,Monkman2026}. We first consider the minimal bilayer system and show that interlayer tunneling substantially enriches the Fermi-surface configurations. In the absence of Rashba spin-orbit coupling (RSOC), the mirror symmetry and subsystem chiral symmetry allow the system to host first-order topological superconducting phases with one or multiple pairs of gapless Majorana edge modes, including helical and additional counterpropagating pairs. These phases are jointly characterized by the mirror Chern number and the winding numbers defined along high-symmetry lines. Upon introducing RSOC, the helical Majorana edge modes are gapped, giving rise to either isolated or twofold Majorana zero modes at each corner. Remarkably, within a certain parameter regime, RSOC gaps out only the helical Majorana pairs while leaving another pair of counterpropagating Majorana edge modes gapless, thereby realizing a hybrid-order topological superconducting phase in which gapless Majorana edge modes coexist with isolated Majorana corner modes. 

Intriguingly, we find that, when all layers are identical, the system possesses a symmetry arising from the layer geometry. 
This layer symmetry allows a decomposition of the bilayer Hamiltonian into two independent sectors with distinct effective chemical potentials. Consequently, the topology of the system can be analyzed via the reduced Hamiltonian and characterized by 
the direct sum of the topological invariants in each sector. Generalizing this construction to an arbitrary $N$-layer system, we find that an analogous layer symmetry can always be defined from the corresponding interlayer-coupling matrix. Moreover, by tuning the chemical potential and the interlayer coupling strengths, one can control how many of these sectors are topologically nontrivial, giving rise to a remarkably rich landscape of topological superconducting phases. These results demonstrate that combining spin-triplet pairing with layer degrees of freedom enables highly tunable topological superconducting phases that are absent in the monolayer limit.

The remainder of this paper is organized as follows. In Sec.~\ref{sec2}, we introduce the bilayer model Hamiltonian and analyze its symmetries. Section \ref{sec3} discusses the Fermi-surface configurations and first-order topological superconducting phases. In Sec.~\ref{sec4}, we investigate the second- and hybrid-order topological superconducting phases. The generalization to arbitrary $N$-layer systems is presented in Sec.~\ref{sec5}. 
Finally, in Sec.~\ref{sec6}, we discuss the symmetry-protection mechanisms and conclude the paper.

\section{Bilayer model Hamiltonian \label{sec2}}
\begin{figure}[htp]
	\centering
	\includegraphics[width=0.5\textwidth]{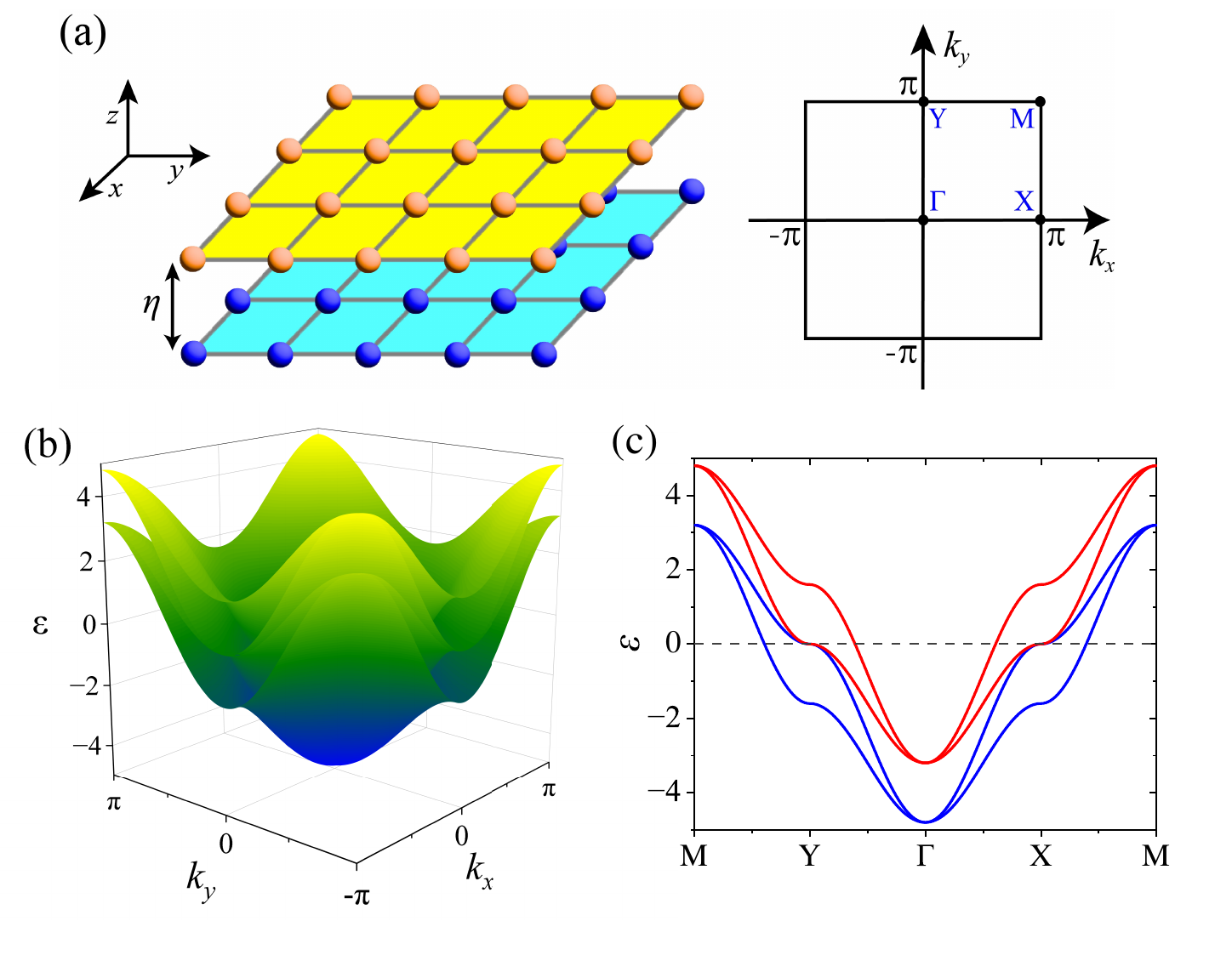}
	\caption{
		(a) Schematic of the bilayer superconducting $d$-wave altermagnetic metal (left) and the corresponding Brillouin zone (right), with the four time-reversal-invariant momenta highlighted. Normal-state energy spectra (b) in the whole Brillouin zone and (c) along high symmetry lines. The red and blue solid lines in (c) correspond to the $\sigma_{x}=1$ and $\sigma_{x}=-1$ sectors, respectively, while the black dashed line indicates the Fermi level. The parameters are $t=1$, $\eta=0.8$, $\lambda_{\rm so}=0$, and $t_{\rm AM}=0.2$.
	}\label{fig1}
\end{figure}

We first focus on the bilayer system composed of two superconducting $d$-wave altermagnetic metal layers, as depicted in Fig.~\ref{fig1}(a). The tight-binding Hamiltonian describing the bilayer system is given by $H_{\rm B}=\frac{1}{2}\sum_{\bk}\Psi_{\bk}^{\dag}\mathcal{H}_{\rm B}(\bk)\Psi_{\bk}$, where the basis function is chosen as
$\Psi^{\dag}_{\bk}=(c_{t}^{\dag}, c_{b}^{\dag})$, with $c_{\alpha}^{\dag}=(c_{\alpha,\uparrow,\bk}^{\dag} ,c_{\alpha,\downarrow,\bk}^{\dag}, c_{\alpha,\uparrow,-\bk}, c_{\alpha,\downarrow,-\bk})$. The Bogoliubov-de Gennes (BdG) Hamiltonian takes the general form
\begin{eqnarray}
	\mathcal{H}_{\rm B}(\bk)=\begin{pmatrix}
	\mathcal{H}_{t}(\bk) & \eta \\
	\eta & \mathcal{H}_{b}(\bk)
	\end{pmatrix},\label{BdGH}
\end{eqnarray}  
where $\mathcal{H}_{t/b}(\bk)$ describes the top/bottom layer superconductor in the decoupled limit, and $\eta$ represents the interlayer coupling strength. In this work, we assume that the top and bottom layers are identical and denote their common monolayer BdG Hamiltonian by $\mathcal{H}_{0}(\bk)$, which is given by~\cite{Zhu2023topological}
\begin{eqnarray}
	\mathcal{H}_{0}(\bk)&=&[-2t(\cos k_{x}+\cos k_{y})-\mu]\tau_{z}  \nonumber \\
	&&+2\lambda_{\rm so}(\sin k_{x}\tau_{z}s_{y}-\sin k_{y}s_{x}) \nonumber \\
	&&+2t_{\rm AM}(\cos k_{x}-\cos k_{y})\tau_{z} s_{z} \nonumber \\
	&&+2 \Delta_p (\sin k_{x} \tau_{x} s_{z}+ \sin k_{y} \tau_{y} ), \label{monoBdGH}
\end{eqnarray}
where the Pauli matrices $\{ \tau_{i}\}$ and $\{ s_{i} \}$ act on the particle-hole and spin ($\uparrow, \downarrow$) degrees of freedom, respectively. The parameter $t$ denotes the strength of the intralayer nearest-neighboring hopping, $\mu$ denotes the chemical potential, and $\lambda_{\rm so}$ denotes the strength of RSOC, which can naturally arise if the altermagnetic metal is grown on a substrate. In line with previous work~\cite{Zhu2023topological,Monkman2026}, we assume helical 
$p$-wave superconductivity. Although the pairing generally contains mixed-parity components when inversion symmetry is broken by the presence of RSOC, we neglect the even-parity spin-singlet components. These components are typically secondary to the spin-triplet component in altermagnets with weak RSOC, and their effects on the topology of the superconducting phase amount to no more than a shift of the phase boundary. For notational simplicity, the lattice constant $a$ is set to unity, and all identity matrices are left implicit throughout.

The BdG Hamiltonian in Eq.~(\ref{BdGH}) can also be expressed explicitly in terms of Pauli matrices as
\begin{eqnarray}
	\mathcal{H}_{\rm B}(\bk)&=&[-2t(\cos k_{x}+\cos k_{y})-\mu]\tau_{z}+\eta \sigma_{x}\tau_{z} \nonumber \\
	&&+2\lambda_{\rm so}(\sin k_{x}\tau_{z}s_{y}-\sin k_{y}s_{x}) \nonumber \\
	&&+2t_{\rm AM}(\cos k_{x}-\cos k_{y})\tau_{z} s_{z} \nonumber \\
	&&+2 \Delta_p (\sin k_{x} \tau_{x} s_{z}+ \sin k_{y} \tau_{y} ). \label{BdGH2}
\end{eqnarray}
The corresponding normal-state Hamiltonian $h(\bk)$, which describes the bilayer $d$-wave altermagnetic metal, is given by
\begin{eqnarray}
	h(\bk)&=&-2t(\cos k_{x}+\cos k_{y}) +\eta \sigma_{x} \nonumber \\
	&&+2\lambda_{\rm so}(\sin k_{x}s_{y}-\sin k_{y}s_{x}) \nonumber \\
	&&+2t_{\rm AM}(\cos k_{x}-\cos k_{y}) s_{z}. \label{normal Hamiltonian}
\end{eqnarray}
Here, the additional Pauli matrices $\{ \sigma_{i} \}$ act on the layer ($t, b$) degree of freedom. From Eq. (\ref{BdGH2}), the pairing matrix $\Delta(\bk)$ in the Nambu space can be extracted as
\begin{eqnarray}
	\Delta(\bk)=2 \Delta_p (\sin k_{x}s_{z}- i \sin k_{y}).
\end{eqnarray}
The pairing matrix satisfies $\mathcal{P}\Delta(\bk)\mathcal{P}^{T}=-\Delta(-\bk)$ with $\mathcal{P}=\sigma_{x}$ being the inversion-symmetry operator of the normal-state Hamiltonian, suggesting the odd-parity nature of the pairing.

The BdG Hamiltion possesses particle-hole symmetry $\Xi=\tau_{x}\mathcal{K}$ ($\mathcal{K}$ is the complex conjugation operator). The $d$-wave altermagnetic exchange field breaks time-reversal symmetry $\mathcal{T}=is_{y}\mathcal{K}$ and fourfold rotation symmetry $\mathcal{C}_{4z}=e^{-i\frac{\pi}{4}\tau_{z}s_{z}}$ individually but preserves their combination $\mathcal{C}_{4z}\mathcal{T}=ie^{-i\frac{\pi}{4}\tau_{z}s_{z}}s_{y}\mathcal{K}$. Intriguingly, the bilayer model additionally possesses a layer symmetry $\mathcal{L}_{\rm B}=\sigma_{x}$, which satisfies the commutation relation with the BdG Hamiltonian $\mathcal{H}_{\rm B}(\bk)$, i.e., $[\mathcal{L}_{\rm B},\mathcal{H}_{\rm B}(\bk)]=0$. Therefore, $\mathcal{H}_{\rm B}(\bk)$ becomes block-diagonal in the basis where $\mathcal{L}_{\rm B}$ takes the diagonal form $\text{diag}\{I_{4\times4},-I_{4\times4}\}$. Here,  $I_{4\times 4}$ is the $4\times4$ identity matrix. The diagonal form of $\mathcal{H}_{\rm B}(\bk)$ can be expressed as $\mathcal{H}_{\rm B}(\bk)=\mathcal{H}_{\rm B}^{+}(\bk)\oplus \mathcal{H}_{\rm B}^{-}(\bk)$, where the superscript $+(-)$ indicates that the block corresponds to the positive (negative) eigenvalue of $\mathcal{L}_{\rm B}$. The Hamiltonians for the two layer-hybridized sectors are given by
\begin{eqnarray}
	\mathcal{H}_{\rm B}^{\pm}(\bk)&=&\mathcal{H}_{0}|_{\mu=\mu_{\rm B}}(\bk), \label{bilayerH}
\end{eqnarray}
where $\mu_{\rm B}=\mu \mp \eta$ denote the effective chemical potentials for the $\mathcal{L}_{\rm B}=\pm 1$ sectors. The bilayer system can thus be regarded as two decoupled monolayer helical $p$-wave superconductors with distinct effective chemical potentials $\mu_{\rm B}$. As demonstrated below, this symmetry can stabilize a number of novel topological superconducting phases.

\section{First-order topological superconductivity \label{sec3}}
\subsection{Fermi-surface configurations}
We first focus on the limit without RSOC, i.e., $\lambda_{\rm so}=0$. In this case, the spin is conserved, and the helical 
$p$-wave superconductor can realize helical topological superconducting phases, which are characterized by the presence of helical Majorana edge states 
at the boundary~\cite{Zhu2023topological}.  As noted earlier, due to the odd-parity nature of the $p$-wave pairing, 
the band topology of the helical topological superconductor is closely tied to the Fermi-surface 
configuration~\cite{Fu2010odd,Sato2010topological}. Therefore, before analyzing the band topology of the 
resulting superconducting phases, we examine how the Fermi-surface configuration evolves with the chemical 
potential and determine how many qualitatively distinct configurations can arise in this bilayer setting.
 
Since the two eigenvalues of $\sigma_{x}$ are good quantum numbers, the energy spectra of the normal-state Hamiltonian in Eq.~(\ref{normal Hamiltonian}) are readily obtained as
\begin{eqnarray}
	\varepsilon_{\alpha \beta}(\bk)&=&-2t(\cos k_{x}+\cos k_{y}) +\alpha \eta \nonumber \\
	&&+2\beta |t_{\rm AM}(\cos k_{x}-\cos k_{y})|
\end{eqnarray}
where the indices $\alpha, \beta=\pm 1$. The above spectra consist of two copies of the two-band spectra of a monolayer altermagnet, shifted relative to each other by $2|\eta|$. An explicit illustration of the band structure is shown  
in Figs.~\ref{fig1}(b) and \ref{fig1}(c), with the model parameters chosen as $\{t,\eta,\lambda_{\rm so},t_{\rm AM}\}=\{1,0.8,0,0.2\}$.
It is noteworthy that  the energy spectra are doubly degenerate throughout the Brillouin zone when $t_{\rm AM}=0$ due to 
spin-SU(2) symmetry. The presence of $d$-wave altermagnetic exchange field lifts this symmetry, producing spin-split bands everywhere except along the $|k_{x}|=|k_{y}|$ lines. This $d$-wave spin splitting gives rise to  Fermi surfaces with opposite spin polarizations that 
are connected by a $C_{4z}$ rotation. Consequently, Fermi surfaces with opposite spin polarizations must appear in pairs, ensuring a zero net magnetization.

\begin{figure*}[htp]
	\centering
	\includegraphics[width=1\textwidth]{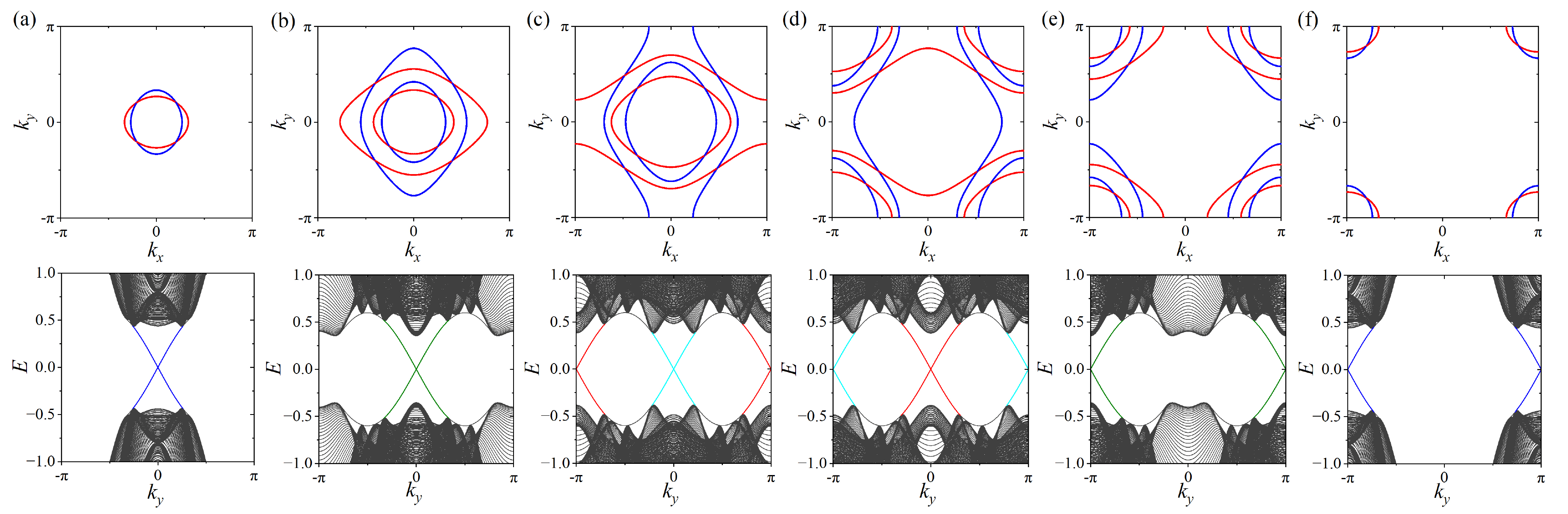}
	\caption{
		Upper panels: Fermi-surface configurations of the normal states. The red and blue colors denote the spin polarizations $\langle s_{z} \rangle=1$ and $-1$ on the Fermi-surface segments, respectively. Lower panels: Energy spectra of the corresponding superconducting phases under open (periodic) boundary conditions in the $x$ ($y$) direction, with pairing amplitude $\Delta_p=0.3$. The red, blue, cyan and green solid lines represent nondegenerate, doubly degenerate, triple degenerate and quadruple degenerate edge states, respectively. (a) $\mu=-4$, (b) $\mu=-2$, (c) $\mu=-1$, (d) $\mu=1$, (e) $\mu=2$, and (f) $\mu=4$.  Common parameters are $t=1$, $\eta=0.8$, $\lambda_{\rm so}=0$, and $t_{\rm AM}=0.2$.
	}\label{fig2}
\end{figure*}

For the band structure shown in Figs.~\ref{fig1}(b) and \ref{fig1}(c),  the corresponding evolution of the Fermi-surface configurations with respect to the chemical potential $\mu$ is illustrated in the upper panels of Fig.~\ref{fig2}. Upon increasing $\mu$ from negative to positive values, the Fermi level first crosses the two lowest bands, giving rise to two elliptical Fermi-surface segments with opposite spin polarizations around the Brillouin-zone center $\mathbf{\Gamma} =(0,0)$ [Fig.~\ref{fig2}(a)]. After the Fermi level crosses the next pair of band edges, a second pair of Fermi-surface segments emerges, resulting in four $\Gamma$-centered segments [Fig.~\ref{fig2}(b)]. As each elliptical segment inflates, the major-axis tips first reach either the $\mathbf{X}=(\pi,0)$ or $\mathbf{Y}=(0,\pi)$ point, depending on its orientation [Figs.~\ref{fig2}(c) and \ref{fig2}(d)]. The segment then crosses the Brillouin-zone boundary and splits into disconnected pieces. Subsequently, the minor-axis tips reach the other high-symmetry point, causing these pieces to reconnect and form a new Fermi-surface segment centered at $\mathbf{M}=(\pi,\pi)$ [Figs.~\ref{fig2}(e) and \ref{fig2}(f)]. This two-step Lifshitz evolution progressively converts the $\mathbf{\Gamma}$-centered Fermi-surface segments into $\mathbf{M}$-centered ones. With a further increase in $\mu$, the $\mathbf{M}$-centered Fermi-surface segments gradually shrink to points and eventually disappear in the Brillouin zone. According to the number of Lifshitz transitions in the Fermi-surface topology, we find that, if we restrict ourselves to the situations hosting Fermi surfaces, there are six distinct configurations. Consequently, we expect six distinct topological superconducting phases.

\subsection{Topological superconducting phases}

When RSOC is absent,  the bilayer superconducting system possesses the mirror symmetry $\mathcal{M}_{z}=i\sigma_{x}s_{z}$. Since $\mathcal{M}_{z}$ commutes with the BdG Hamiltonian, i.e., $[\mathcal{M}_{z}, \mathcal{H}_{\rm B}(\bk)]=0$, the Hamiltonian  $\mathcal{H}_{\rm B}(\bk)$ can be decoupled as $\mathcal{H}_{\rm B}(\bk)=\mathcal{H}_{\rm B}^{+i}(\bk)\oplus \mathcal{H}_{\rm B}^{-i}(\bk)$ in the basis where $\mathcal{M}_{z}$ takes the diagonal form $\text{diag}\{ iI_{4\times 4}, -iI_{4\times 4} \}$.  The Hamiltonians of the two mirror sectors are given by
\begin{eqnarray}
	\mathcal{H}_{\rm B}^{\pm i}(\bk)&=&[-2t(\cos k_{x}+\cos k_{y})-\mu]\tau_{z}  \nonumber \\
	&&+[\eta \pm 2t_{\rm AM}(\cos k_{x}-\cos k_{y}) ]\tau_{z} \rho_{z} \nonumber \\ 
	&&\pm 2\Delta_{p}\sin k_{x}\tau_{x}\rho_{z} +2\Delta_{p}\sin k_{y}\tau_{y},
	\label{mirrorH}
\end{eqnarray}
where $\{ \rho_{i}\}$ denotes the Pauli matrices acting on the degrees of freedom formed by hybridizing the layer and spin degrees of freedom. One readily finds that both mirror-sector Hamiltonians retain particle-hole symmetry, while time-reversal symmetry and chiral symmetry are broken. According to the conventional tenfold-way classification~\cite{Schnyder2008classification,Kitaev2009periodic,Ryu2010topological}, these two Hamiltonians belong to symmetry class D and are therefore characterized by Chern numbers~\cite{Ueno2013symmetry}
\begin{eqnarray}
	C_{\pm i}=\frac{1}{2\pi}\sum_{n}\int_{BZ}\Omega_{\pm i}^{(n)}(\bk) d\bk,
\end{eqnarray} 
where $\Omega_{\pm i}^{(n)}(\bk)$ represents the Berry curvature of the $n$-th negative-energy band for the $\pm i$ mirror-sector. Since the full Hamiltonian respects $C_{4z}\mathcal{T}$ symmetry, the total Chern number is constrained to vanish identically~\cite{Zhu2023topological}. The Chern numbers for the two mirror sectors thus satisfy $C_{+i}=-C_{-i}$. This allows us to define the mirror Chern number $C_{M}=(C_{+i}-C_{-i})/2$ to characterize the first-order topology of the system~\cite{Teo2008surface}. The $Z$-valued $C_{M}$ permits multiple pairs of helical Majorana modes 
stably coexist on the edge. Here, the precise meaning of `helical' refers to a pair of counterpropagating edge states 
whose dispersions satisfy $E_{s}(k_{y})=E_{\bar{s}}(-k_{y})$ with $s$ labeling the chirality, analogous to the case with time-reversal symmetry.

We numerically calculate the BdG spectra corresponding to the Fermi-surface configurations in Fig.~\ref{fig2} under open (periodic) boundary conditions in the $x$ ($y$) direction. A direct comparison reveals that the number of zero-energy spectral crossings at $k_{y}=0$ or $k_{y}=\pi$ exactly matches the number of intersections between the Fermi-surface segments and the respective high-symmetry lines in the Brillouin zone, confirming 
that the topology of the superconducting phase is directly related to the underlying Fermi-surface configuration. Specifically, for the topological superconducting phases shown in Figs.~\ref{fig2}(a) and \ref{fig2}(f), the system hosts a pair of helical Majorana modes per edge, 
which is consistent with the calculated mirror Chern number $C_{M}=-1$. This result indicates that each mirror sector carries one chiral edge mode, and 
the chiral edge modes in the two mirror sectors have opposite chirality. This means that the system realizes  a topological mirror superconductor characterized by a pair of helical Majorana edge modes~\cite{Zhang2013topological}. For the topological superconducting phases shown in Figs.~\ref{fig2}(b) and \ref{fig2}(e), the system carries two pairs of helical Majorana edge modes. However, we find $C_{M}=0$. This suggests that each mirror sector 
hosts a pair of helical Majorana edge modes. Since time-reversal symmetry is broken, the stability of these helical Majorana modes is protected
by subsystem symmetry, as we will demonstrate below. Furthermore, for the topological superconducting phases shown in Figs.~\ref{fig2}(c) and \ref{fig2}(d), the system hosts four branches of gapless Majorana modes on each edge. Together with the calculated mirror Chern number $C_{M}=1$, this indicates that one mirror sector carries three chiral Majorana branches, one of which has chirality opposite to the other two. The other mirror sector carries a single Majorana branch whose chirality is opposite to the net chirality of the former mirror sector. The absence of hybridization between these chiral Majorana modes with opposite chirality is also due to subsystem symmetry. 

In Fig.~\ref{fig2}, the numerical results show that all gapless Majorana edge modes cross at the high-symmetry points 
$k_{y}=0$ or $k_{y}=\pi$ in the boundary Brillouin zone. This spectral property implies 
that additional weak topological invariants, defined along the high-symmetry lines $k_{y}=0$ or $k_{y}=\pi$ of the 
bulk Brillouin zone, enables a refined classification 
of the band topology of the superconducting phases. 
Along these two high-symmetry lines,  the BdG Hamiltonian $\mathcal{H}_{\rm B}(\bk)$ reduces to the effective one-dimensional form:
\begin{eqnarray}
	\mathcal{H}_{\rm B}(k_{x},0/\pi )&=&[-2t(\cos k_{x} \pm 1)-\mu]\tau_{z} +\eta \sigma_{x}\tau_{z} \nonumber \\
	&&+2t_{\rm AM}(\cos k_{x}\mp 1)\tau_{z} s_{z} \nonumber \\
	&&+2 \Delta_p \sin k_{x} \tau_{x} s_{z},
\end{eqnarray}
where the upper and lower signs correspond to the lines $k_{y}=0$ and $k_{y}=\pi$, respectively. It is readily verified that both reduced one-dimensional Hamiltonians exhibit an emergent chiral symmetry $\mathcal{S}=\tau_{y}s_{z}$, satisfying $\{\mathcal{S},\mathcal{H}_{\rm B}(k_{x},0/\pi )\}=0$. This chiral symmetry allows winding numbers to be defined for each reduced Hamiltonian, whose explicit forms are given by~\cite{Ryu2010topological}
\begin{eqnarray}
	W_{0/\pi}&=&\frac{1}{4\pi i}\oint dk_{x} {\rm{Tr}} [\mathcal{S} \mathcal{H}_{\rm B}^{-1}(k_{x}, 0/\pi)  \nonumber \\
	&&\times \partial_{k_{x}}\mathcal{H}_{\rm B}(k_{x}, 0/\pi)].
	\label{wn}
\end{eqnarray}
A straightforward calculation reveals that the winding-number pairs for the superconducting phases shown in Fig.~\ref{fig2} are $(W_{0}, W_{\pi})=(-2,0), (-4,0), (-3,-1), (-1,-3), (0,-4)$, and $(0,-2)$, respectively. The values of the winding numbers are in exact agreement with the number of zero-energy states per boundary at $k_y=0$ and $k_y=\pi$.

\subsection{Topological phases diagram}
\begin{figure}[htp]
	\centering
	\includegraphics[width=0.50\textwidth]{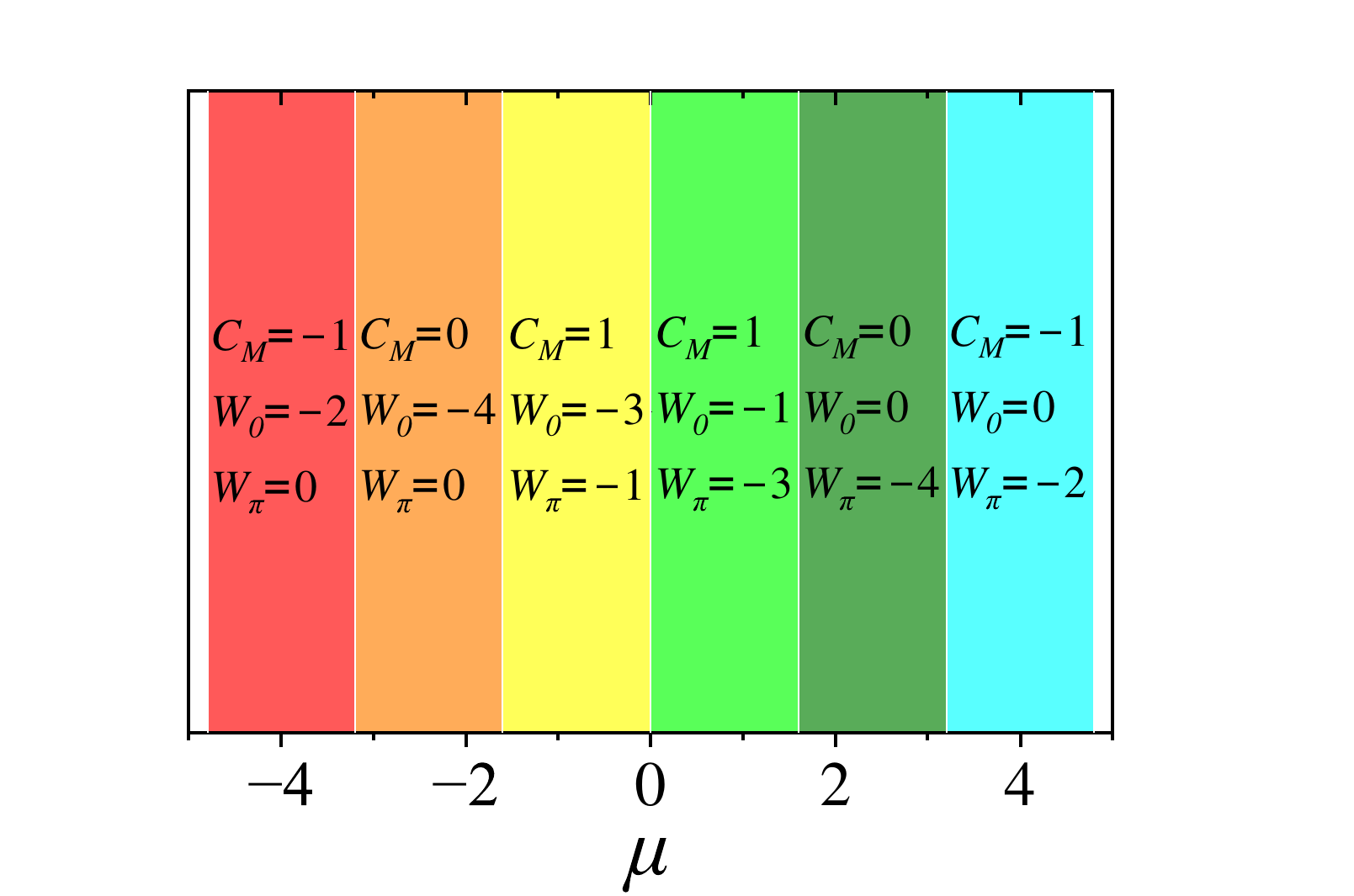}
	\caption{
		Topological phase diagram as a function of $\mu$. Parameters are chosen as $t=1$, $\eta=0.8$, $\lambda_{\rm so}=0$, $\Delta_p=0.3$ and $t_{\rm AM}=0.2$.
	}\label{fig3}
\end{figure}

The mirror Chern number $C_{M}$, together with the winding numbers $W_{0/\pi}$, provides a comprehensive characterization of the first-order topological superconducting phases. Therefore, we determine the complete phase diagram as a function of $\mu$ in terms of these topological invariants. Since the change of first-order topology necessarily involves the closing and reopening of the bulk gap, the phase boundaries can be determined from the gap-closing conditions. From the BdG Hamiltonian in Eq.~(\ref{BdGH2}), a straightforward analysis reveals that the gap closings occur only at the time-reversal-invariant momenta $\mathbf{K}= \{\mathbf{\Gamma}, \mathbf{X}, \mathbf{Y}, \mathbf{M} \}$, where the superconducting pairing terms vanish. Accordingly, the energy spectra at these momenta take the following form
\begin{eqnarray}
	E_{\alpha \beta \gamma}(\mathbf{K})&=&	\alpha|-2t(\cos K_{x}+\cos K_{y})-\mu  +\beta \eta  \nonumber \\
	&&+2\gamma t_{\rm AM}(\cos K_{x}-\cos K_{y})|, 
\end{eqnarray} 
where $\alpha, \beta, \gamma=\pm 1$. Therefore, the gap closes when the chemical potential satisfies the relations
\begin{eqnarray}
	\mu=\left\{\begin{array}{cc}
		-4t+\beta \eta, & \mathbf{\Gamma}=(0,0), \\
		\beta \eta -4\gamma t_{\rm AM}, & \mathbf{X}=(\pi,0), \\
		\beta \eta +4\gamma t_{\rm AM}, & \mathbf{Y}=(0,\pi), \\
		4t+\beta \eta, & \mathbf{M}=(\pi,\pi).
	\end{array}\right.
\end{eqnarray}
It is worth noting that the energy-gap closings at the $\mathbf{X}$ and $\mathbf{Y}$ points occur simultaneously due to 
$\mathcal{C}_{4z}\mathcal{T}$ symmetry. 
We numerically verify our analytical derivation and obtain the phase diagram characterized by $C_{M}$ and $W_{0/\pi}$, 
as shown in Fig.~\ref{fig3}. As the chemical potential $\mu$ varies, the system exhibits six distinct topologically nontrivial regimes, corresponding to the phases illustrated in Fig.~\ref{fig2}.

\section{Second-order and hybrid-order topological superconductivity \label{sec4}}

The helical Majorana edge modes in the first-order topological superconducting phases are protected by mirror symmetry $\mathcal{M}_{z}$
and subsystem chiral symmetry $\mathcal{S}$.  Both symmetries are lifted once RSOC becomes finite. 
In the monolayer system, Zhu {\it et al.} previously demonstrated that RSOC-induced lifting of these symmetries gaps out the helical Majorana edge modes and induces opposite Dirac masses on two adjacent $\mathcal{C}_{4z}$-related edges, thereby leading to the appearance of Majorana zero modes per corner (also dubbed Majorana corner modes for brevity)~\cite{Zhu2023topological}. According to the Jackiw-Rebbi theory~\cite{Jackiw1976solitons}, gapping out a pair of helical Majorana edge modes can leave behind a single Majorana zero mode at each Dirac-mass domain wall~\cite{Zhu2018tunable,Langbehn2017reflection}. Without additional symmetry protection, Majorana corner modes follow a $Z_{2}$ classification~\cite{Schnyder2008classification,Kitaev2009periodic,Ryu2010topological}. One would therefore naively expect that, upon coupling two identical copies of the system, two Majorana zero modes at the same corner hybridize and split away from zero energy. Such splitting can be avoided only in the presence of some emergent symmetries that extend the classification from $Z_{2}$ to $Z$, with chiral symmetry typically playing this role~\cite{Mo2025topological,Yin2025multifold,Zhu2024Direct}. In this work, however, we find that the combination of odd-parity pairing and the layer degree of freedom naturally provides such a protecting symmetry.

\begin{figure}[htp]
	\centering
	\includegraphics[width=0.5\textwidth]{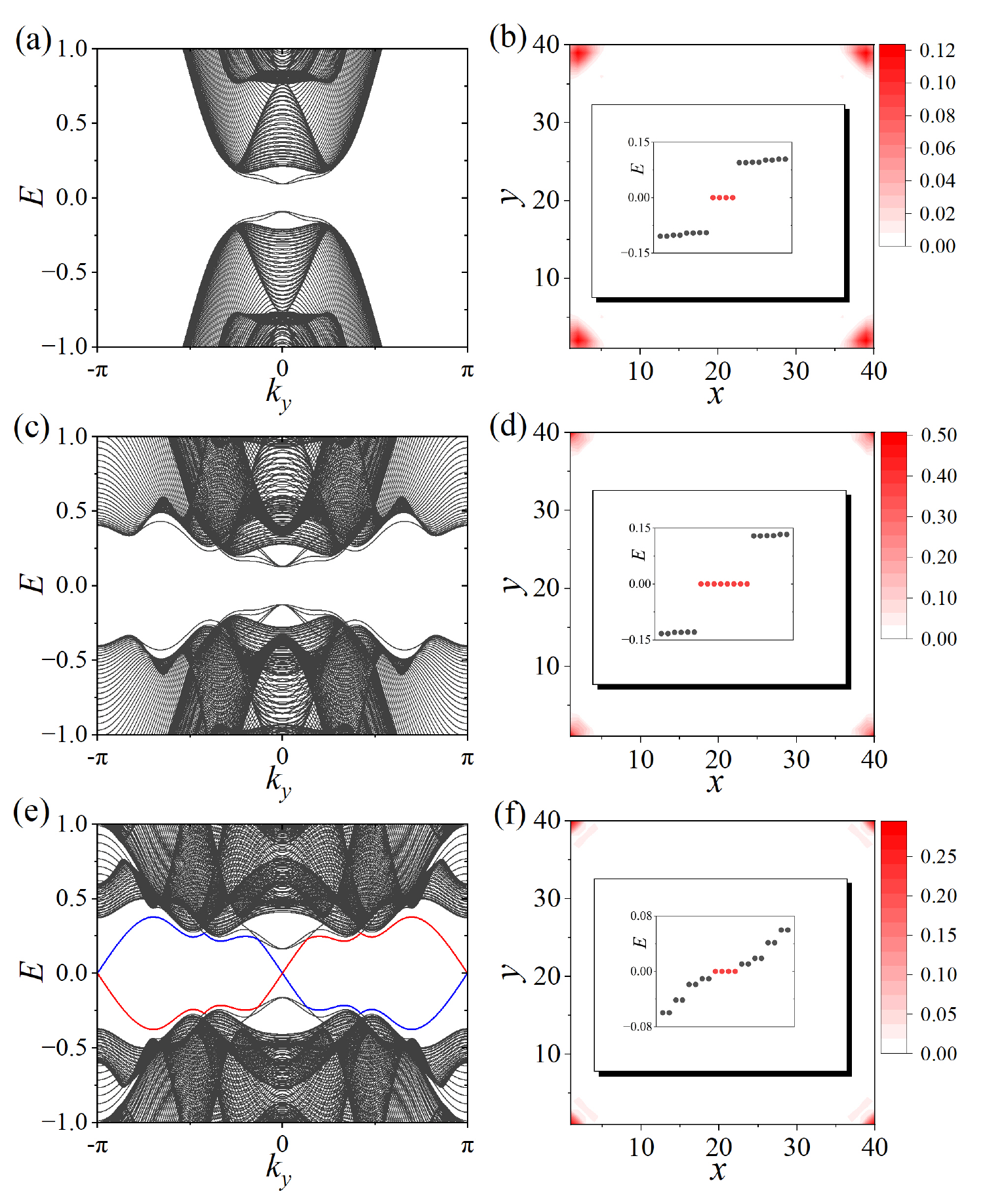}
	\caption{
		Left panels: Energy spectra for the bilayer system under open (periodic) boundary conditions in the $x$ ($y$) direction. Right panels: Probability density profiles of Majorana corner modes for the system under open boundary conditions in both the $x$ and $y$ directions. The red and blue solid lines in (e) represent the midgap states on the left and right edges, respectively. In (a) and (b) $\mu=-4$, in (c) and (d) $\mu=-2$, in (e) and (f) $\mu=-1$. Common parameters are $t=1$, $\eta=0.8$, $\lambda_{\rm so}=0.2$, $\Delta_p=0.3$, and $t_{\rm AM}=0.2$.
	}\label{fig4}
\end{figure}

In the present bilayer system, finite RSOC is expected to likewise gap out the helical Majorana edge modes, and the richer edge-state patterns compared with the monolayer case enable the emergence of a broader variety of second-order topological
superconducting phases. Here, we consider a finite but weak RSOC ($\lambda_{\rm so}\ll t$). Figure \ref{fig4} 
shows the energy spectra obtained by including RSOC in the first-order topological superconducting phases of Figs.~\ref{fig2}(a)--\ref{fig2}(c).
As expected, all helical Majorana edge modes 
 are gapped out, giving rise to isolated or twofold Majorana zero modes at each corner of the square geometric system. 
The number of Majorana zero modes per corner equals the number of helical Majorana edge modes
that become gapped. Interestingly, 
for the topological superconducting phase shown in Fig.~\ref{fig2}(c), RSOC gaps out only the pair of helical Majorana edge modes crossing at $k_{y}=0$, thereby generating a single Majorana zero mode per corner (isolated Majorana corner modes). Meanwhile, the other pair of counterpropagating Majorana edge modes, comprising one branch crossing zero energy at $k_{y}=0$ and the other at $k_y=\pi$, remains robust against RSOC. As a result, the system exhibits gapless Majorana edge modes in the ribbon geometry and zero-energy Majorana corner modes in the square geometry, as shown in Figs.~\ref{fig4}(e) and \ref{fig4}(f), respectively. This coexistence of nontrivial first- and second-order topological boundary states is the defining characteristic of the hybrid-order topological phase~\cite{Wu2023hybrid}. It is worth noting that the topological superconducting phases for $\mu>0$, corresponding to Figs.~\ref{fig2}(d)–\ref{fig2}(f), exhibit exactly the same behaviors and are therefore not shown here.

To understand why not all Majorana edge modes are gapped out in the hybrid-order topological superconducting phase, we note that, in addition to the subsystem chiral symmetry $\mathcal{S}$ discussed above, an additional chiral symmetry exists along the high-symmetry lines
$k_{y}=0$ and $k_{y}=\pi$ when RSOC is absent. This symmetry, characterized by
$\tilde{\mathcal{S}}=\tau_{y}$, arises as the product of  $\mathcal{S}$ and the spin-U(1) symmetry $U_{s}=s_{z}$ that 
appears in this limit. Although a finite RSOC breaks both $\mathcal{S}$ and the spin-U(1)
symmetry, their product $\tilde{\mathcal{S}}$ survives. We can therefore define winding numbers
 $\tilde{W}_{0/\pi}$ following 
the same procedure as in Eq.~(\ref{wn}).
A direct calculation yields $(\tilde{W}_{0}, \tilde{W}_{\pi})=(1,1)$ for the hybrid-order phase shown in Fig.~\ref{fig4}(e), 
in agreement with the numbers of Majorana zero modes per edge at these two momenta. 

\begin{figure}[htp]
	\centering
	\includegraphics[width=0.5\textwidth]{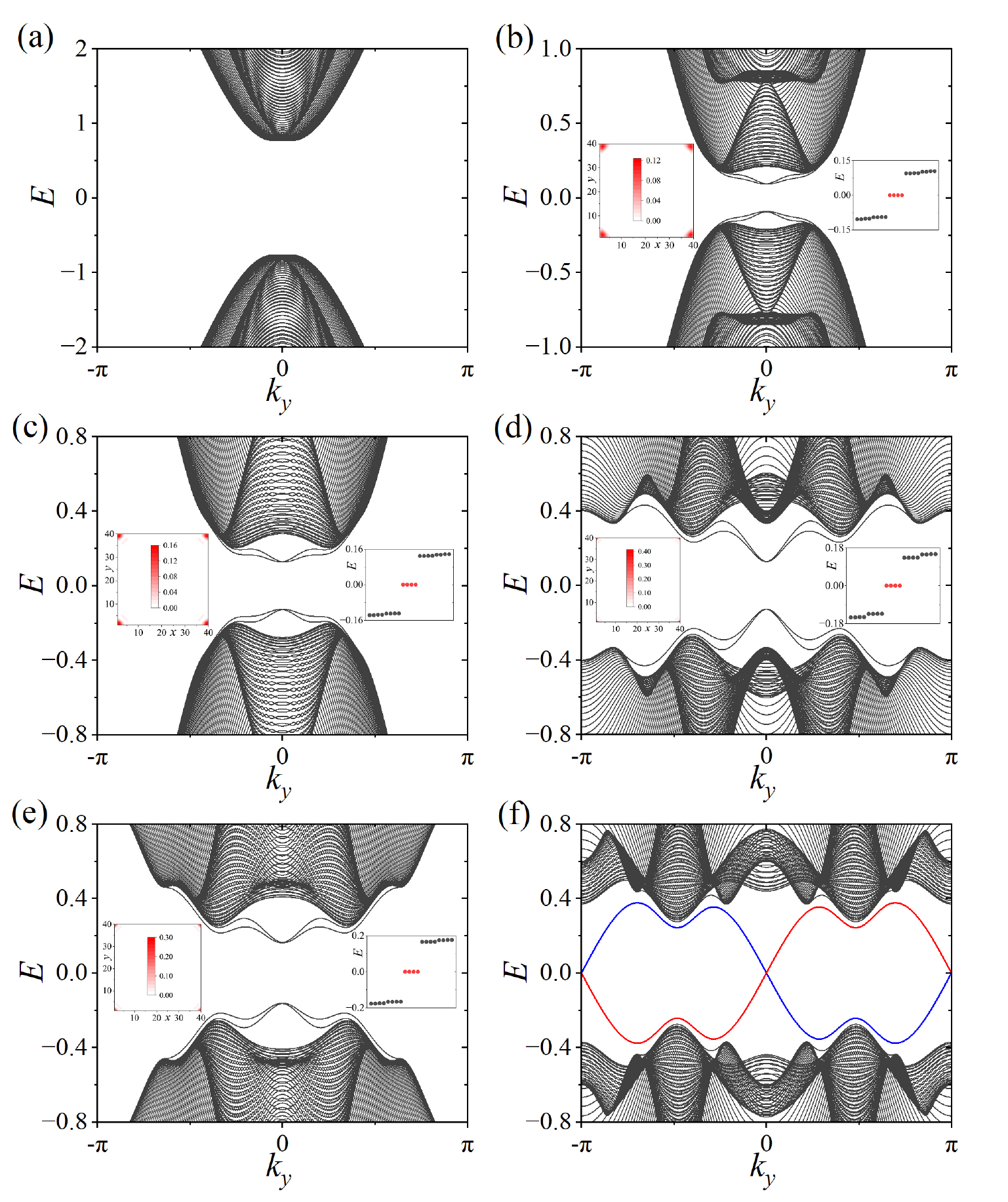}
	\caption{Energy spectra under open (periodic) boundary conditions in the $x$ ($y$) direction. Left panels:  $\mathcal{L}_{\rm B}=1$ sector. Right panels:  $\mathcal{L}_{\rm B}=-1$ sector. Insets show the probability density profiles of the Majorana corner modes in the $\mathcal{L}_{\rm B}$ sectors when both $x$ and $y$ directions take open boundary conditions. The red and blue solid lines in (f) represent the midgap states on the left and right edges, respectively. In (a) and (b) $\mu=-4$, in (c) and (d) $\mu=-2$, in (e) and (f) $\mu=-1$. Common parameters are $t=1$, $\eta=0.8$, $\lambda_{\rm so}=0.2$, $\Delta_p=0.3$, and $t_{\rm AM}=0.2$.
	}\label{fig5}
\end{figure}

As discussed earlier, the bilayer model possesses a layer symmetry $\mathcal{L}_{\rm B}$. This symmetry allows the full BdG 
Hamiltonian in Eq.~(\ref{BdGH2}) to be decomposed into two independent sectors labeled by its eigenvalues. Despite the strong interlayer coupling, each sector is described by a monolayer-like Hamiltonian with a distinct effective chemical potential $\mu_{\rm B}$, as given in Eq.~(\ref{bilayerH}). We next numerically calculate the energy spectra of the two $\mathcal{L}_{\rm B}$ sectors for the three phases shown in Fig.~\ref{fig4}, with the results presented in Fig.~\ref{fig5}. For the second-order topological superconducting phase with isolated Majorana corner modes, the $\mathcal{L}_{\rm B}=1$ sector is topologically trivial, whereas the $\mathcal{L}_{\rm B}=-1$ sector hosts a single Majorana zero mode at each corner [see Figs.~\ref{fig5}(a) and \ref{fig5}(b)]. For the phase with twofold Majorana corner modes, each $\mathcal{L}_{\rm B}$ sector contributes one Majorana zero mode at each corner [see Figs.~\ref{fig5}(c) and \ref{fig5}(d)]. By contrast, in the hybrid-order topological superconducting phase, the $\mathcal{L}_{\rm B}=1$ sector realizes a second-order topological superconducting phase, while the $\mathcal{L}_{\rm B}=-1$ sector lies in a weak first-order topological superconducting phase [see Figs.~\ref{fig5}(e) and \ref{fig5}(f)]. These results provide an intuitive physical picture for the origin of the topological superconducting phases shown in Fig.~\ref{fig4}.

\section{Generalization to $N$-layer systems \label{sec5}}
The bilayer system can be straightforwardly generalized to an arbitrary number of layers. When all layers are identical, 
the $N$-layer counterpart ($N\geq 2$) is described by the BdG Hamiltonian
\begin{eqnarray}
	\mathcal{H}_{\rm N}(\bk)&=&[-2t(\cos k_{x}+\cos k_{y})-\mu]\tau_{z}+ \zeta \tau_{z}   \nonumber \\
	&&+2\lambda_{\rm so}(\sin k_{x}\tau_{z}s_{y}-\sin k_{y}s_{x}) \nonumber \\
	&&+2t_{\rm AM}(\cos k_{x}-\cos k_{y})\tau_{z} s_{z} \nonumber \\
	&&+2 \Delta_p (\sin k_{x} \tau_{x} s_{z}+ \sin k_{y} \tau_{y} ), 
\end{eqnarray}
where $\zeta$ denotes the $N \times N$ interlayer-coupling matrix, and the identity matrix acting on the layer degree of freedom is omitted for convenience. 

\begin{figure}[htp]
	\centering
	\includegraphics[width=0.48\textwidth]{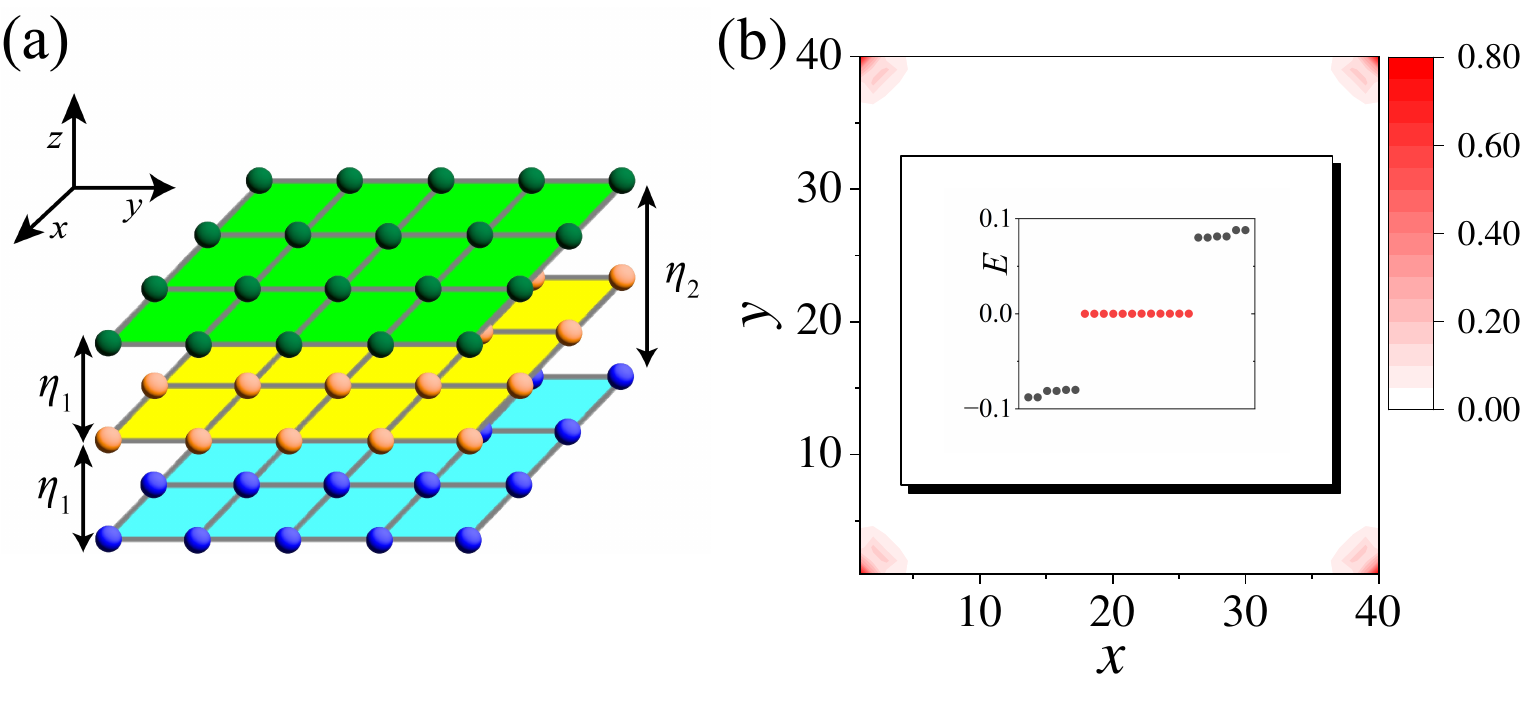}
	\caption{
	(a) Schematic of the trilayer system. (b) Probability density profiles of the $12$ Majorana zero modes (three per corner) for the trilayer system under open boundary conditions in both the $x$ and $y$ directions. The parameters are $t=1$, $\mu=-2$, $\eta_{1}=0.8$, $\eta_{2}=0.4$, $\lambda_{\rm so}=0.2$, $\Delta_p=0.3$, and $t_{\rm AM}=0.2$.
	}\label{fig6}
\end{figure}

Without loss of generality, we next focus on the trilayer case ($N=3$) depicted in Fig.~\ref{fig6}(a). To avoid confusion, we denote the BdG Hamiltonian by $\mathcal{H}_{\rm T}(\bk)$ and the coupling matrix among the three layers ($t,m,b$) by $\tilde{\zeta}$, with
\begin{eqnarray}
	\tilde{\zeta}=\begin{pmatrix}
	0 & \eta_1 & \eta_2  \\
	\eta_1 & 0 & \eta_1  \\
	\eta_2 & \eta_1 & 0 \end{pmatrix}.
\end{eqnarray}
Here, $\eta_1$ and $\eta_2$ denotes the nearest- and next-nearest- layer coupling strengths, respectively. The basis function is chosen as $	\tilde{\Psi}_{\bk}^{\dag}=(c_{t}^{\dag} , c_{m}^{\dag}, c_{b}^{\dag})$ with $c_{\alpha}^{\dag}=(c_{\alpha,\uparrow,\bk}^{\dag} ,c_{\alpha,\downarrow,\bk}^{\dag}, c_{\alpha,\uparrow,-\bk}, c_{\alpha,\downarrow,-\bk})$.

Increasing the number of layers introduces additional internal degrees of freedom, suggesting that the trilayer system may host an even richer variety of topological superconducting phases. Notably, this trilayer system also preserves a layer symmetry $\mathcal{L}_{\rm T}=\tilde{\zeta}$, satisfying $[\mathcal{L}_{\rm T}, \mathcal{H}_{\rm T}(\bk)]=0$. Similar to the bilayer case, the BdG Hamiltonian $\mathcal{H}_{\rm T}(\bk)$ becomes block-diagonal in the basis where $\mathcal{L}_{\rm T}$ takes the diagonal form $\text{diag}\{\tilde{\zeta}_{1}I_{4\times4},\tilde{\zeta}_{2}I_{4\times4}, \tilde{\zeta}_{3}I_{4\times4}\}$. Here, $\tilde{\zeta}_1$, $\tilde{\zeta}_{2}$ and $\tilde{\zeta}_{3}$ are the the eigenvalues of $\tilde{\zeta}$. Direct diagonalization yields $\tilde{\zeta}_{1}=-\eta_{2}$, $\tilde{\zeta}_{2}=(\eta_{2}-\sqrt{8\eta_{1}^{2}+\eta_{2}^{2}})/2$, and $\tilde{\zeta}_{3}=(\eta_{2}+\sqrt{8\eta_{1}^{2}+\eta_{2}^{2}})/2$. Accordingly, $\mathcal{H}_{\rm T}(\bk)$ can be decomposed into three independent sectors and written as $\mathcal{H}_{\rm T}(\bk)=\mathcal{H}_{\rm T}^{\tilde{\zeta}_{1}}(\bk)\oplus \mathcal{H}_{\rm T}^{\tilde{\zeta}_{2}}(\bk) \oplus \mathcal{H}_{\rm T}^{\tilde{\zeta}_{3}}(\bk)$, where 
\begin{eqnarray}
	\mathcal{H}_{\rm T}^{\tilde{\zeta}_{i}}(\bk)= \mathcal{H}_{0}|_{\mu=\mu_{\rm T}}(\bk). \label{trilayerH}
\end{eqnarray}
Here, $\mu_{\rm T}=\mu-\tilde{\zeta}_{i} (i=1,2,3)$ denotes the effective chemical potentials for the three $\mathcal{L}_{\rm T}$ sectors. Evidently, in each sector, the Hamiltonian describes an effective monolayer model with a distinct effective chemical potential $\mu_{\rm T}$. Therefore, by tuning the chemical potential $\mu$ and the interlayer couplings $\eta_{1}$ and $\eta_{2}$, the effective chemical potentials of all three sectors can be simultaneously brought into the topologically nontrivial regime. The trilayer system can thus host an even larger number of helical Majorana edge modes in the absence of RSOC, or higher-fold Majorana corner modes in its presence. As shown in Fig.~\ref{fig6}(b), when all three sectors are tuned into the second-order topological regime, the trilayer system indeed realizes threefold Majorana corner modes.

More generally, for arbitrary $N$, one can always identify a layer symmetry operator $\mathcal{L}_{\rm N}=\zeta$ that satisfies $[\mathcal{L}_{\rm N}, \mathcal{H}_{\rm N}(\bk)]=0$. Accordingly, the full Hamiltonian can be block-diagonalized as
\begin{eqnarray}
	\mathcal{H}_{\rm N}(\bk)=\mathcal{H}_{\rm N}^{\zeta_{1}}(\bk)\oplus \mathcal{H}_{\rm N}^{\zeta_{2}}(\bk) \oplus \cdots \oplus \mathcal{H}_{\rm N}^{\zeta_{N}}(\bk), 
\end{eqnarray}
where $\{ \zeta_{1}, \zeta_{2},\cdots, \zeta_{N} \}$ denotes the set of eigenvalues of $\zeta$. The Hamiltonian in each sector takes the same form as those in Eqs.~(\ref{bilayerH}) and (\ref{trilayerH}) and is characterized by an effective chemical potential $\mu_{\rm N}=\mu-\zeta_{i} (i=1,2,\cdots,N)$. By tuning the effective chemical potential $\mu_{\rm N}$, the number of topologically nontrivial sectors can be controlled in principle. This platform is therefore particularly appealing for the tunability it offers in controlling the number of Majorana modes.

\section{Discussions and conclusions \label{sec6}}

So far, we have assumed that all layers are identical. In realistic materials, however, inhomogeneity 
in the chemical potential, RSOC, and pairing amplitude across layers may exist. It is therefore instructive to examine the 
impact of such inhomogeneity on the stability of the topological superconducting phases discussed in 
this work. We focus on weak inhomogeneity, whose effect on phase stability depends on whether it breaks 
or preserves the protecting symmetry. The symmetries relevant to phase stability in this work include 
particle-hole symmetry, mirror symmetry, layer symmetry, and subsystem chiral symmetry. Particle-hole 
symmetry is intrinsic to the BdG formalism and therefore immune to inhomogeneity; it provides natural 
protection for isolated Majorana corner modes, so that the second-order topological superconducting phase 
hosting them is stable against inhomogeneity. For the remaining three symmetries, we again separate the 
discussion into cases with and without RSOC.

We first focus on the case without RSOC. 
When different layers carry distinct chemical potential (each layer is assumed to be homogeneous, so that $k_{x}$ and $k_{y}$ remain 
good quantum numbers), the mirror symmetry $\mathcal{M}_{z}$ is generally broken. However, the layer symmetry and subsystem chiral symmetry 
remain preserved. This follows from two observations. (i) The Hamiltonian can be decomposed into two parts, $\mathcal{H}=\mathcal{H}_{1}+\mathcal{H}_{2}$, where $\mathcal{H}_{1}$ contains the interlayer coupling terms and the layer-dependent chemical potential, while $\mathcal{H}_{2}$ collects the remaining layer-independent terms. We then define $\tilde{\mathcal{H}}_{1}\equiv \mathcal{H}_{1}l_{0}\tau_{z}s_{0}$, where $l_{0}$ and $s_{0}$ denote the identity matrices in the layer and spin subspace, respectively. 
Since $[\tilde{\mathcal{H}}_{1},\mathcal{H}]=0$, the Hamiltonian is block-diagonalized in the eigenbasis of the momentum-independent matrix $\tilde{\mathcal{H}}_{1}$,
so the layer symmetry survives. (ii) The subsystem chiral symmetry operators, both $\mathcal{S}$ and $\tilde{\mathcal{S}}$, 
are layer-independent. These symmetries suffice to protect the first-order topological superconducting phases 
discussed in this work. When the pairing amplitude becomes layer-dependent, the layer symmetry is broken, since 
no constant matrix acting only on the layer degree of freedom commutes with the BdG Hamiltonian. Nevertheless, 
the subsystem chiral symmetry remains intact, which by itself suffices to protect the spectral crossings of 
the helical Majorana edge modes.

In the presence of RSOC, the mirror symmetry $\mathcal{M}_{z}$ is broken, 
and the layer symmetry also ceases to be well defined once the RSOC and pairing amplitudes 
become layer dependent. Nevertheless, the subsystem chiral symmetry  
$\tilde{\mathcal{S}}$ remains preserved. Together with particle-hole symmetry, it protects 
 the hybrid-order phase, in which isolated Majorana corner modes coexist with counterpropagating Majorana edge states. 
 Owing to the loss of layer symmetry under such inhomogeneity, multifold Majorana corner modes are generally not robust and undergo energy splitting. Despite appearing as a negative result, this symmetry-based conclusion remains instructive: it suggests strategies for seeking suitable materials or designing systems where emergent symmetries can protect such exotic topological superconducting phases.

To conclude, we have shown that layered $d$-wave altermagnetic metals with odd-parity helical $p$-wave pairing can host a rich variety of topological superconducting phases, enabled by the subsystem symmetry and layer symmetry arising from the layer structure. Experimentally, these phases can be probed by scanning tunneling spectroscopy through the spatial distribution of zero-bias signals at the corners and along the edges. Looking ahead, we note that layered systems can exhibit rich nonsymmorphic symmetries depending on the lattice structure of each layer and the stacking configuration. Our findings, together with this observation, suggest that layered altermagnetic metals provide a fertile ground for exploring the emergence of novel topological superconducting phases from the interplay of symmetry and unconventional pairing.

\section{Acknowledgements}
This work is supported by the Fundamental and Interdisciplinary Disciplines Breakthrough Plan of the Ministry of Education of China (Grant No. JYB2025XDXM403) and the Guangdong Basic and Applied Basic Research Foundation (Grant No. 2023B1515040023).

\appendix

\bibliography{AM_helical}

\end{document}